\documentclass{article}

\usepackage[nonatbib,preprint]{neurips_data_2023}

\usepackage[utf8]{inputenc} 
\usepackage[T1]{fontenc}    
\usepackage{url}            
\usepackage{booktabs}       
\usepackage{amsfonts}       
\usepackage{nicefrac}       
\usepackage{microtype}      
\usepackage{xcolor}         
\usepackage[colorlinks,citecolor=blue]{hyperref} 
\usepackage[capitalize,nameinlink]{cleveref}
\usepackage[maxbibnames=99]{biblatex}
\usepackage{xspace}
\usepackage{adjustbox}
\usepackage{multirow}
\usepackage{caption}
\usepackage{subcaption}
\usepackage{graphicx}
\usepackage{wrapfig} 
\usepackage[linesnumbered,ruled,vlined]{algorithm2e}
\newcommand{\name}{\textsc{Context-NER}\xspace}

\title{\name{}: Contextual Phrase Generation at Scale}

\author
{Himanshu Gupta$^{1}$  \hspace{9pt}  Shreyas Verma$^{2}$ \hspace{9pt}  Santosh Mashetty$^{1}$  \hspace{9pt} Swaroop Mishra$^{1}$ \hspace{9pt}\\
\small{$^{1}$School of Computing and AI, Arizona State University}\\
\small{$^{2}$College of Computing, Georgia Institute of Technology} 
\\ \tt\small hgupta35@asu.edu, \tt\small shreyas.verma@gatech.edu\\
}
\begin{document}

\maketitle

\begin{abstract}

Named Entity Recognition (NER) has seen significant progress in recent years, with numerous state-of-the-art (SOTA) models achieving high performance. 
However, very few studies have focused on the generation of entities' context.
In this paper, we introduce \name{}, a task that aims to generate the relevant context for entities in a sentence, where the context is a phrase describing the entity but not necessarily present in the sentence. 
To facilitate research in this task, we also present the EDGAR10-Q dataset, which consists of annual and quarterly reports from the top 1500 publicly traded companies. 
The dataset is the largest of its kind, containing 1M sentences, 2.8M entities, and an average of 35 tokens per sentence, making it a challenging dataset. 
We propose a baseline approach that combines a phrase generation algorithm with inferencing using a 220M language model, achieving a ROUGE-L score of 27\% on the test split.
Additionally, we perform a one-shot inference with ChatGPT, which obtains a 30\% ROUGE-L, highlighting the difficulty of the dataset. 
We also evaluate models such as T5 and BART, which achieve a maximum ROUGE-L  of 49\% after supervised finetuning on EDGAR10-Q. 
We also find that T5-large, when pre-finetuned on EDGAR10-Q, achieve SOTA results on downstream finance tasks such as Headline, FPB, and FiQA SA, outperforming vanilla version by 10.81 points.
To our surprise, this 66x smaller pre-finetuned model also surpasses the finance-specific LLM BloombergGPT-50B by 15 points. 
We hope that our dataset and generated artifacts will encourage further research in this direction, leading to the development of more sophisticated language models for financial text analysis
\footnote{Dataset, the script to generate it, baseline approach, ChatGPT evaluations, and finetuned models are freely available at \url{https://github.com/him1411/edgar10q-dataset}}. \end{abstract}


\section{Introduction}

Recent advancements in Named Entity Recognition (NER) have led to impressive results through the development of various large-scale pretrained models \cite{Zhang2023PromptNERAP,Ma2023CoLaDaAC,Zhang2022OptimizingBF,yuan-etal-2021-improving,wang-etal-2021-improving,wang-etal-2021-automated}. 
While existing research has primarily focused on NER task performance \cite{zhang2022finbert,7009718,francis2019transfer,alexander2021research,wu2022robust,wang2022sentence,varshney-etal-2022-commonsense,shrimal-etal-2022-ner}, limited attention has been given to exploring contextual information associated with the identified entities within sentences. 
To illustrate this challenge, consider the sentence "The rent due today is \$500." 
In this example, the question "What is \$500?" can be answered as "rent due today." 
This task becomes particularly challenging when sentences are lengthy, lack explicit contextual cues, or contain multiple entities. 
For instance, referring to the first sentence in Table \ref{tab1}, discerning the context of "29.2 Million" as "Carrying amount of loans 90 days or more past due" solely based on the sentence becomes difficult. 
Notably, the exact phrase is not directly present in the sentence; however, it holds significant relevance in accurately describing the entity.

\begin{table*}[t!]
\centering
\resizebox{\linewidth}{!}
{
\begin{tabular}{c|ll|l}
\hline
\textbf{Sentences} &
  \textbf{Entity} &
  \textbf{Type} &
  \textbf{Context of entity} \\ \hline
\begin{tabular}[c]{@{}l@{}}As of June 30, 2019, the department store loans \\ discussed above were 90 days or greater past due, \\ as were \$4.5 million of residential loans and a \\ \$36.2 million infrastructure loan with a carrying \\ value of \$29.2 million, net of a \$7.0 million \\ non accretable difference.\end{tabular} &
  \$29.2 Million &
  Money &
  \begin{tabular}[c]{@{}l@{}}Carrying amount of \\ loans 90 days or \\ more past due\end{tabular} \\ \hline
\begin{tabular}[c]{@{}l@{}}There were impairments of \$0.8 million for \\ the three months ended June 30, 2020 and \$2.2 \\ million for the six months ended June 30, 2020.\end{tabular} &
  \$2.2 million &
  Money &
  \begin{tabular}[c]{@{}l@{}}Valuation allowance \\ for loan servicing rights \\ with impairments\end{tabular} \\ \hline
\end{tabular}
}
\caption{Illustration of some example sentences where the entites' relevant context (Column 4) is difficult to generate but relevant for describing the entity (Column 2). 
} 
\label{tab1}
\end{table*}

We propose \name{}, a task which involves generating a relevant phrase that describing an entity within a given sentence, regardless of whether the phrase is explicitly present in the sentence. 
We also introduce the EDGAR10-Q dataset \footnote{ Named after the Electronic Data Gathering, Analysis, and Retrieval system, which performs automated collection, validation, indexing, acceptance, and forwarding of submissions by companies and others who are required by law to file forms with the U.S. Securities and Exchange Commission (SEC)} to initiate a systematic study of the task.
The dataset comprises of quarterly and annual financial reports from publicly traded limited liability companies (LLCs).
These reports are prepared by domain experts (financial analysts), ensuring highest quality of gold labels and contextual relevance. 
Table \ref{tab1} shows some examples of the dataset from which it is evident that phrases are not always present in sentences and can be difficult to retrieve without adequate knowledge of the domain. 
EDGAR10-Q is one of the largest in the financial domain (1M sentences 2.78M entites) and consists of complex sentences that are not prevalent in benchmark datasets, posing a new challenge for SOTA models. 
The dataset has two unique qualities that make it particularly challenging.
Firstly, the sentences are long and complex in nature; averaging approximately 35 tokens per sentence, surpassing the length of sentences typically encountered in the training of large language models (LLMs).
Secondly, since this dataset is prepared from financial documents, they contain several numerical entities whose context can be difficult to extract by using just one sentence (\S \ref{sec:dataset}). 

\begin{wraptable}{l}{0.5\textwidth}
\centering
\fontsize{9pt}{\baselineskip}\selectfont 
\renewcommand\tabcolsep{1pt} 
\renewcommand\arraystretch{1} 
\resizebox{\linewidth}{!}
{
\begin{tabular}{c|c}
\hline
\textbf{Baseline Response} & \textbf{ChatGPT Response}                                                                                    \\ \hline
infrastructure loan        & Infrastructure loan carrying value                                                                           \\ \hline
Impairment        & \begin{tabular}[c]{@{}l@{}}Impairment expenses for three and \\ six months ended June 30, 2020.\end{tabular} \\ \hline
\end{tabular}
}
\caption{Responses of the baseline approach and ChatGPT for sentences present in Table \ref{tab1} highlighting the difficulty to generate relevant context associated with entites. } 
\label{tab2}
\end{wraptable} 


We conduct various experiments in algorithmic (rule based), one-shot and supervised learning settings. 
We introduce a baseline method that leverages syntactic trees of the sentence to generate questions and find relevant phrases in the sentences (\S \ref{sec:approach}). 
The baseline yields an overall result of 27.59 ROUGE-L score . 
We also conduct a one shot evaluation on ChatGPT \cite{brown2020language} to get 30.31\% score. 
Responses of the baseline approach and ChatGPT are illustrated in Table \ref{tab2}.
We also train T5's \cite{2020t5} different variants (T5, Tk-Instruct\cite{wang-etal-2022-super}, and Flan T5\cite{Chung2022ScalingIL}) and BART model \cite{lewis2020bart} in a supervised manner to get 49\% as the highest ROUGE-L score (\S \ref{results}).
The low scores are identified as an area for further research to enhance the learning capabilities for such complex tasks. 


We examine the effects of the generated artifacts using the dataset. 
Our findings reveal that T5 pre-finetuned on EDGAR10-Q outperforms vanilla T5 by 10.81 points and surpasses BloombergGPT 50B by 15.81 points on various downstream finance datasets (\S \ref{downstream_tasks}). 
Additionally, we provide a comparison between \name{} and OpenIE to highlight the distinctions between these tasks. 
Lastly, we explore the effect of instruction tuning on the T5 model using the EDGAR10-Q dataset, resulting in a performance improvement of 2\% points.

\noindent\textbf{Contributions:} 
(a) we introduce the task \name{}, to generate contextual phrases for entities in sentences and associated EDGAR10-Q dataset created from financial reports;
(c) we evaluate the dataset using the following methods:
(c.1) we introduce a baseline approach which achieves a 27\% ROUGE-L;
(c.2) we evaluate the dataset in a one-shot setting via ChatGPT achieving 30\% ROUGE-L;
(c.3) we train different generative models in a supervised manner to get $\sim50\%$ performance;
(d) we perform a detailed analysis on following lines of enquiry
(d.1) effect of pre-finetuning using EDGAR10-Q to achieve SOTA on several finance downstream tasks
(d.2) qualitative comparison of \name{} with OpenIE
(d.3) explore the effect of instruction tuning for EDGAR10-Q.

\section{EDGAR10-Q Dataset}
\label{sec:dataset}

\begin{figure}[h!] 
\begin{minipage}{0.48\textwidth} 
    \centering
    \centering
    \fontsize{8.5pt}{\baselineskip}\selectfont 
    \resizebox{5 cm}{!}
    {
        \begin{tabular}{l|l}
        \hline
        \multicolumn{1}{c|}{\textbf{Entity Types}}                                        & \multicolumn{1}{c}{\textbf{Counts}} \\ \hline
        \begin{tabular}[c]{@{}l@{}}Floating Values\\ (monetary and percent)\end{tabular}  & 2143054                             \\
        \begin{tabular}[c]{@{}l@{}}number of Assets \\ (Shares and Integers)\end{tabular} & 425850                              \\
        Ordinal Values                                                                    & 16891                               \\
        Dates                                                                             & 195174                               \\ \hline
        \end{tabular}
    }
    \captionof{table}{Entity-wise statistics that show the number of entities for each category.}
    \label{tab3}
    \end{minipage}
 \hfill
 \begin{minipage}{0.48\textwidth} 
    \centering
    \fontsize{8.5pt}{\baselineskip}\selectfont 
    {\color{black}
    \resizebox{5 cm}{!}
    {
        \begin{tabular}{c|c}
        \hline
        \textbf{Other Statistics} & \textbf{Values} \\ \hline
        Entites per sentence      & 1.78           \\
        Words per paragraph       & 113.14           \\
        Labels per entity         & 1.45           \\
        Words per sentence        & 35.88           \\ \hline
        \end{tabular}
    }}
    \captionof{table}{Other relevant statistics of the dataset highlighting its size and difficulty. }
    \label{tab4}
 \end{minipage}

\end{figure}
\begin{figure}[h!] 
\begin{minipage}{0.45\textwidth} 
    \centering
    \centering
    \fontsize{8.5pt}{\baselineskip}\selectfont 
    \resizebox{7.5 cm}{!}
    {
    \begin{tabular}{l|llll}
    \hline
    \textbf{Dataset name} & \textbf{Docs} & \textbf{Sentences} & \textbf{Words} & \textbf{Entities} \\ \hline
    funsd\cite{jaume2019} & 200 & NA & 31485 & 9743 \\
    wikicoref\cite{Ghaddar2016WikiCorefAE} & 30 & 2229 & 59652 & 3557 \\
    scierc\cite{luan2018multitask} & 500 & NA & NA & 8089 \\
    med ment.\cite{Patil_2020} & 4392 & 42602 & 1176058 & 352496 \\
    genia\cite{Fu2020NestedNE} & 2000 & 18545 & 436967 & 96582 \\
    conll 2003\cite{Sang2003IntroductionTT} & 1393 & 22137 & 301418 & 35089 \\ \hline
    \textbf{EDGAR10-Q} & \textbf{18752} & \textbf{1009712} & \textbf{77400425} & \textbf{2780969} \\ \hline
    \end{tabular}
    }
    \captionof{table}{Comparison of EDGAR10-Q with other NER datasets}
    \label{tab5}
    \end{minipage}
 \hfill
 \begin{minipage}{0.51\textwidth} 
    \centering
    \fontsize{8.5pt}{\baselineskip}\selectfont 
    {\color{black}
    \resizebox{5.5 cm}{!}
    {
        \begin{tabular}{l|ll|ll}
        \hline
         & \multicolumn{2}{l|}{\textbf{Train}} & \multicolumn{2}{l}{\textbf{Test}} \\ \cline{2-5} 
        \textbf{\begin{tabular}[c]{@{}l@{}}\# of\\ Entities\end{tabular}} & \textbf{\begin{tabular}[c]{@{}l@{}}\# of\\ Sent.\end{tabular}} & \textbf{\begin{tabular}[c]{@{}l@{}}Avg.\\ S. Len\end{tabular}} & \textbf{\begin{tabular}[c]{@{}l@{}}\# of\\ Sent.\end{tabular}} & \textbf{\begin{tabular}[c]{@{}l@{}}Avg.\\ S. Len\end{tabular}} \\ \hline
        \textbf{1} & 381251 & 31.40 & 53567 & 30.26 \\
        \textbf{2} & 261687 & 35.82 & 31985 & 35.18 \\
        \textbf{3} & 86020 & 41.97 & 10425 & 41.93 \\
        \textbf{4} & 38500 & 52.02 & 4309 & 54.23 \\
        \textbf{5 +} & 16726 & 73.01 & 1530 & 85.98 \\ \hline
        \textbf{Overall} & \textbf{784184} & \textbf{35.93} & \textbf{101816} & \textbf{34.85} \\ \hline
        \end{tabular}
    }
    }
    \captionof{table}{Train and test split of the dataset for the supervised learning.}
    \label{tab6}
 \end{minipage}

\end{figure}

\paragraph{Dataset Creation:} 

The EDGAR10-Q dataset was created by scraping quarterly (10-Q) and annual (10-K) reports from the years 2019, 2020, and 2021. 
Given the crucial role these meticulously prepared reports play in assessing the financial health of organizations, great care is taken in their curation to ensure accuracy and quality, leaving no room for oversight.
To ensure standardization, all SEC filings undergo a tagging process where entities within sentences are labeled with corresponding Named Entity Recognition (NER) context labels, serving as high-quality gold labels for the dataset. 
We refer the reader to \S \ref{creation} for more details. 


\paragraph{Dataset Description:}
Table \ref{tab3} shows the four types of entities, namely money, time (duration), percent, and cardinal values (pure, shares, and integer) present in the data. 
Table \ref{tab4} further elucidate data richness through paragraph and sentence level statistics. 
Table \ref{tab5} compares this dataset with benchmark NER datasets and contains nearly 18.7K documents comprising 1M sentences. 
As it can be inferred from the table, the EDGAR-10Q dataset has nearly 650x more documents and nearly 780x more entities in comparison to the popular NER benchmark dataset - wikicoref \cite{Ghaddar2016WikiCorefAE}. 
We observe that the EDGAR10-Q is the largest and richest in multiple parameters and a first-of-its-kind dataset in the financial domain.
Table \ref{tab6} highlights the train test split of the dataset.
The dataset is also divided according to the different number of entities present in a sentence. 
We see that as the number of entities increases, the average sentence length increase as well. 
Since this is  a real-world dataset, sentences with 1 entity are most prevalent and comprise 49\% of the dataset while sentences with 5+ entities consist of 2\% of the dataset.
The train and test set are of roughly equal difficulty in terms of sentences (\S \ref{datacard}).

\section{Baseline Approach }
\label{sec:approach}

\begin{figure*}[h!]
	\centering
	\includegraphics[width=13cm, height= 4.5 cm]{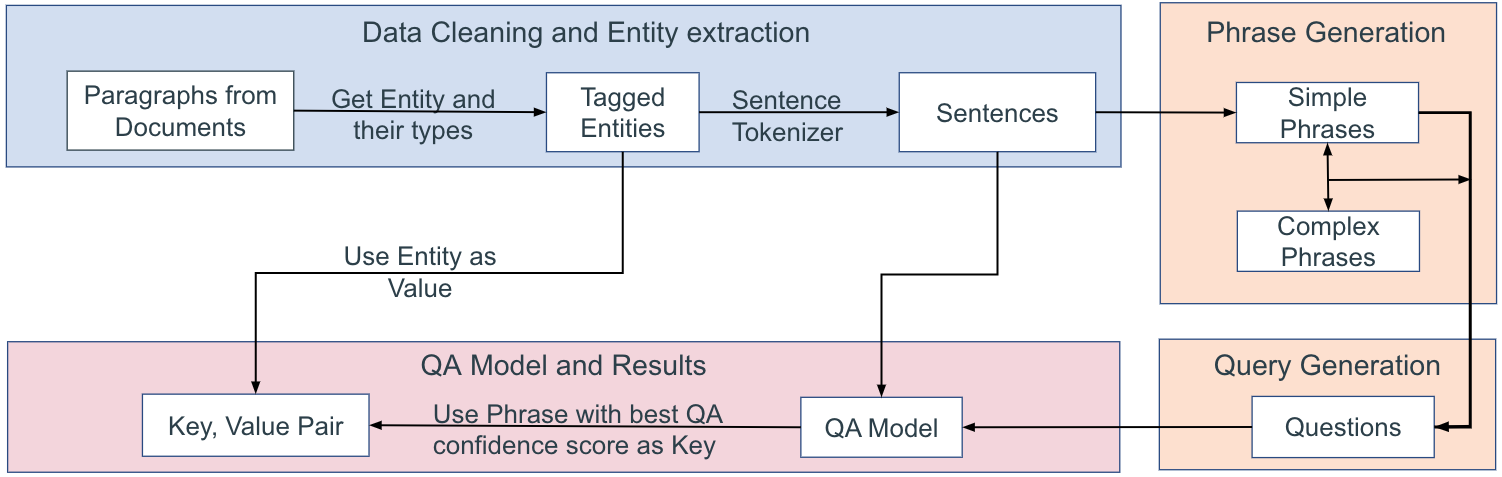}
	\caption{Illustrates the overall process flow of the proposed zero-shot open information extraction technique using question generation and reading comprehension.}
	\label{fig:block}
\end{figure*}

We present a simple, yet efficient method to extract entities' descriptions from sentences, as shown in Figure \ref{fig:block}. 
OpenIE is predicated on the idea that the relation (which is action verbs in most cases) is the central element of the extraction process, from which all other considerations flow.
However, in many cases, the verb is not helpful, particularly in financial data. 
Consider the sentence: "Deferred revenue for 2020 is \$20 billion." 
Like most financial records are of the form "is, was, be," etc., the verb "is" in this sentence is an auxiliary verb and does not describe any particular event or give any information about the entity. 
Our approach consists of a phrase generation algorithm that is in turn used for the creation of questions and fed into a transformer model for machine reading comprehension. 

\begin{wrapfigure}{l}{0.55\textwidth}
  \begin{algorithm}[H]
    \SetStartEndCondition{ }{}{}%
    \SetKwProg{Fn}{def}{\string:}{}
    \SetKwFunction{Range}{range}
    \SetKw{KwTo}{in}\SetKwFor{For}{for}{\string:}{}%
    \SetKwIF{If}{ElseIf}{Else}{if}{:}{elif}{else:}{}%
    \AlgoDontDisplayBlockMarkers
    \SetAlgoNoEnd
    \SetAlgoNoLine%
    \SetKwInput{KwInput}{Input}                
    \SetKwInput{KwOutput}{Output}              
    \DontPrintSemicolon
    \KwInput{Sentence}
    \KwOutput{List of Phrases}
    \SetKwFunction{FMain}{complex\_noun\_phrase\_extractor}
    \SetKwFunction{FSum}{noun\_phrase}
    \SetKwProg{Fn}{Function}{:}{\KwRet}
    \Fn{\FSum{Sentence}}
    {
        doc = sequence\_of\_token(Sentence) \;
        phrase\_list = []    \;
        \For{token in Doc }
        {
            ph = \textquotesingle \hspace{.25cm}  \textquotesingle  \;   
            \If{
            token.head.pos in [Noun, Pronoun] and token.dep in [Object, Subject]  
            }
            {
                \For{subtoken in token.children }
                {
                    \If{subtoken.pos is Adj or subtoken.dep is Comp }
                    {
                        ph += subtoken.text + \textquotesingle \hspace{.15cm} \textquotesingle
                    }
                }
                \If{len(ph) is not 0}
                {
                    ph += token.text
                }
            }
            \If{len(ph) is not 0 and ph doesnot have entities}
            {
                phrase\_list.append(ph)
            }
        }
        \KwRet phrase\_list\;
  }
    \caption{Phrase Generation}
    \label{algo2}
  \end{algorithm}
\end{wrapfigure}

\subsection{Phrase generation }
A noun phrase (NP) is defined as phrase \cite{stuart2013importance} containing a noun, a person, place, or thing, and the modifier that distinguishes it.
We extract two types of phrases from the sentences, namely simple and complex. 
In simple phrase extraction, each sentence comprises subject-object and verb connecting them where the subject or object is usually a noun or pronoun. 
After searching for a noun and pronoun, we check for any noun compound or adjective. 
On the other hand, for complex phrase extraction, we first start with preposition extraction. 
It has to be noted that simple phrases are not always found on both sides of the proposition. 
We then follow similar steps as in simple phrase extraction to look for phrases in both the left and right of the preposition. 
Consider an example,\textit{In connection with the refinance we reduced the loan amount by \$6.8 million.}. 
From our algorithm, the phrases extracted from this sentence are \textit{loan amount} and \textit{connection with refinance}.
We use Spacy \footnote{Spacy POS Tagging Library link: \url{https://spacy.io/usage/linguistic-features}.} library for POS tags of the word which were leveraged in Algorithm \ref{algo2}. 
We refer the reader to \S \ref{baseline} where the approach is described in more detail along with examples and flowchart (Algo. \ref{algo1} and Fig. \ref{syntactc_tree}). 

\subsection{Machine Reading Comprehension for Baseline}
Phrases and entities are extracted on a sentence level for each paragraph. 
Based on the type of entity and the noun phrases, the questions are framed accordingly in a rule-based fashion. 
For instance, if the entity found out was of type date, then the question would be "when is" + NP?. 
Once these questions are generated, they are fed into the MRC Model, and the  answers are checked for the relevant entity. 
If there are multiple questions with the same answer, we select the one with the highest confidence score.
There are instances where none of the generated questions returned an answer with the target entity or returned responses with a different entity. 
For those cases, we create the question "what is" entity? and its response would be considered as the relevant phrase. 
In case these questions return different entities as responses, all cases to identify the noun phrase fail and the algorithm does not return a response.
A detailed description of the MRC model is present in \S \ref{mrc_model}. 
Since this method does not require any finetuning, the baseline is directly evaluated on the test set.

\subsection{ChatGPT Evaluation}
We also establish a ChatGPT baseline which evaluates the test set in a one-shot setting. We provide the definition and the example as shown below: 

\noindent
\fbox{
\begin{minipage}{\linewidth}
{ 
   \textbf{\newline Definition:} Based on the example given below, Given an entity and a sentence containing the entity, generate a phrase that describes the entity in the sentence.
   \newline
   \textbf{Example 1 Input: }\$15.Issuance of common stock in May 2019 public offering at \$243.00 per share, net of issuance costs of \$15.
   \newline
   \textbf{Example 1 Output: }Common stock public offering issuance costs 
}
\end{minipage}
} 

 Like the previous baseline, this method does not require any finetuning, it is directly evaluated on the test set as well.

\section{Experiments and Results}
\subsection{Experimental Setup}
\label{experimental_setup}

\textbf{Baseline Model Setup:} We run all our experiments using the BERT base model \cite{devlin2018bert}. All experiments are done with Nvidia V100 16GB GPU.

\textbf{ChatGPT Setup:} We evaluate ChatGPT (gpt-3.5-turbo, max tokens = 256, top p = 1, frequency penalty = 0, presence penalty = 0) in one-shot setting.

\textbf{Performance Evaluation metrics:} ROUGE-L score uses the longest common subsequence matching between the baseline and GPT-3 responses to compare output quality. 
We report precision, recall, and the F1 measure against the ROUGE-L \cite{lin2004rouge} score. 
We also report the Exact Match \cite{rajpurkar2016squad} which measures the ratio of the instances for which a model's response has a ROUGE-l score of 1 with a gold label.
We report No match where the generated output and gold label 0 ROUGE-L score.

\subsection{Results}
\label{results}

\begin{figure}[t!] 
\begin{minipage}{0.42\textwidth} 
    \centering
    \centering
    \fontsize{8.5pt}{\baselineskip}\selectfont 
    \resizebox{4.5 cm}{!}
    {
        \begin{tabular}{l|lll}
        \hline
        \multirow{2}{*}{\textbf{\begin{tabular}[c]{@{}l@{}}\# of\\ Ent.\end{tabular}}} & \multicolumn{3}{c}{\textbf{Baseline Scores}}     \\
                    & \textbf{P} & \textbf{R} & \textbf{F1} \\ \hline
        \textbf{1}  & 36.69      & 27.04      & 28.19       \\
        \textbf{2}  & 36.57      & 29.53      & 29.66       \\
        \textbf{3}  & 32.67      & 25.97      & 26.48       \\
        \textbf{4}  & 29.59      & 24.11      & 24.27       \\
        \textbf{5+} & 23.82      & 19.35      & 19.56       \\ \hline
        \textbf{Overall}                                                               & \textbf{34.59} & \textbf{27.06} & \textbf{27.59} \\ \hline
        \end{tabular}
    }
    \captionof{table}{Score of Baseline approach on the test set showing precision, recall and F1}
    \label{tab7}
    \end{minipage}
 \hfill
 \begin{minipage}{0.53\textwidth} 
    \centering
    \fontsize{8.5pt}{\baselineskip}\selectfont 
    {\color{black}
    \resizebox{6 cm}{!}
    {
        \begin{tabular}{lll|lll}
        \hline
        \multirow{2}{*}{\textbf{\begin{tabular}[c]{@{}l@{}}\# of \\ Ent.\end{tabular}}} & \multirow{2}{*}{\textbf{\begin{tabular}[c]{@{}l@{}}\# of\\ Sent.\end{tabular}}} & \multirow{2}{*}{\textbf{\begin{tabular}[c]{@{}l@{}}Avg. S. \\ Len.\end{tabular}}} & \multicolumn{3}{c}{\textbf{Chat GPT Scores}} \\
         &  &  & \textbf{P} & \textbf{R} & \textbf{F1} \\ \hline
        1 & 38919 & 29.68 & 25.13 & 42.19 & 28.55 \\
        2 & 24717 & 33.61 & 26.25 & 50.99 & 31.53 \\
        3 & 8131 & 39.03 & 26.60 & 48.86 & 31.54 \\
        4 & 3341 & 50.28 & 26.56 & 50.37 & 31.62 \\
        5+ & 1050 & 75.99 & 20.47 & 43.53 & 25.02 \\ \hline
        \textbf{Overall} & \textbf{76158} & \textbf{33.49} & \textbf{25.72} & \textbf{47.49} & \textbf{30.31} \\ \hline
        \end{tabular}
    }}
    \captionof{table}{Details of ChatGPT performance on a smaller test set. The table shows the smaller test's statistics and ChatGPT's precision, recall, and F1. }
    \label{tab8}
 \end{minipage}

\end{figure}

\paragraph{Baseline Scores:} Table \ref{tab7} shows the baseline results where the overall F1 27.59\%. 
Precision uniformly decreases from 36\% to 23\% while recall is ranging from 19\% to 30\%, leading to an overall F1 score in the range from 20\% to 30\% for each category of the baseline model.
F1 shows a linearly decreasing trend with an increase in the number of entities (the exception being of 2 entity sentences higher than 1). 
Table \ref{tab8} shows ChatGPT scores which is evaluated on a subset of the actual Test set. 
This evaluation set is roughly 75\% of the actual test and is reduced due to budget limitations. 
However, the distribution of this eval set is similar to the actual set. 
Contrasting to the baseline approach, ChatGPT's F1 score stays constant at 31\% and decreases sharply as the number of entities increases to 5+. 
The overall precision of ChatGPT is lower than baseline, but recall is much higher than baseline, resulting in a higher overall F1.
The performance difference between the two increases as the length of sentences increases.
Although the F1 of ChatGPT is higher than the baseline score (30.31\% vs. 27.59\%), there is significant room for improvement.


\begin{table*}[t!]
\resizebox{\linewidth}{!}
{
\begin{tabular}{l|lll|lll|lll|lll|lll}
\hline
\multirow{2}{*}{\textbf{\begin{tabular}[c]{@{}l@{}}\# of \\ Ent.\end{tabular}}} &
  \multicolumn{3}{l|}{\textbf{BART Base}} &
  \multicolumn{3}{l|}{\textbf{T5 Base}} &
  \multicolumn{3}{l|}{\textbf{T5 Large}} &
  \multicolumn{3}{l|}{\textbf{Flan T5 Base}} &
  \multicolumn{3}{l}{\textbf{Tk-Inst Large}} \\ \cline{2-16} 
 &
  \textbf{P} &
  \textbf{R} &
  \textbf{F1} &
  \textbf{P} &
  \textbf{R} &
  \textbf{F1} &
  \textbf{P} &
  \textbf{R} &
  \textbf{F1} &
  \textbf{P} &
  \textbf{R} &
  \textbf{F1} &
  \textbf{P} &
  \textbf{R} &
  \textbf{F1} \\ \hline
1 &
  50.26 &
  45.98 &
  45.71 &
  49.33 &
  44.91 &
  44.50 &
  50.44 &
  46.38 &
  45.76 &
  49.46 &
  45.09 &
  44.65 &
  49.44 &
  45.18 &
  44.66 \\
2 &
  54.44 &
  51.03 &
  50.47 &
  53.65 &
  50.30 &
  49.45 &
  54.96 &
  51.89 &
  50.94 &
  53.96 &
  50.49 &
  49.71 &
  53.81 &
  50.36 &
  49.54 \\
3 &
  54.74 &
  51.56 &
  50.89 &
  53.31 &
  49.84 &
  48.95 &
  55.16 &
  51.82 &
  50.93 &
  53.74 &
  49.97 &
  49.31 &
  53.71 &
  50.01 &
  49.30 \\
4 &
  54.37 &
  51.36 &
  50.69 &
  53.86 &
  50.29 &
  49.49 &
  55.47 &
  52.60 &
  51.66 &
  53.55 &
  50.25 &
  49.37 &
  53.60 &
  50.29 &
  49.42 \\
5+ &
  53.31 &
  50.13 &
  49.63 &
  51.52 &
  47.27 &
  46.97 &
  52.66 &
  48.48 &
  48.24 &
  49.44 &
  45.29 &
  45.12 &
  49.61 &
  45.15 &
  45.02 \\ \hline
\textbf{Overall} &
  53.10 &
  49.51 &
  \textbf{49.01} &
  52.13 &
  48.35 &
  \textbf{47.67} &
  53.49 &
  50.02 &
  \textbf{49.23} &
  52.22 &
  48.40 &
  \textbf{47.77} &
  52.17 &
  48.38 &
  \textbf{47.70} \\ \hline
\end{tabular}
}
\captionof{table}{Supervised learning scores of different models. P, R and F1 denote Precision, Recall and F1 respectively. }
\label{tab9}
\end{table*}

\paragraph{Supervised training results:} Table \ref{tab9} shows the results when the generative models are finetuned with the train split and evaluated on test split. 
The overall supervised training performance is much higher as compared to baselines, where base models (T5, Flan, and BART Base) perform much better than ChatGPT (47.67\%, 47.77\%, and 49.01\% respectively vs. 30.31\%). 
We see that there is no significant improvement in performance as the number of parameters increases. T5 and Tk-Instruct Large (49.23\% and 47.70\%) give nearly the same F1 scores as BART (49.01\%).

\paragraph{Exact Match:} The results of ChatGPT and Baseline are consistently low, as shown in Table \ref{tab10}.
We infer this is because of complex hidden contexts and the sentence structures of the dataset.
ChatGPT's score is consistently lower than the baseline as the recall of ChatGPT is consistently higher due to which obtaining an exact match is difficult. 
Table \ref{tab11} shows the results of supervised learning where consistently higher scores are obtained. 
Instruction-tuned variants of T5 (Flan and Tk-Instruct) perform the best out of the models but the overall score is still low. 

\begin{figure}[t!] 
\begin{minipage}{0.30\textwidth} 
    \centering
    \centering
    \fontsize{9pt}{\baselineskip}\selectfont 
    {\color{black}\resizebox{\textwidth}{!}{
        \begin{tabular}{lll}
        \hline
        \multicolumn{1}{c}{\textbf{\begin{tabular}[c]{@{}c@{}}\# of\\ Entities\end{tabular}}} & \textbf{Baseline} & \textbf{\begin{tabular}[c]{@{}l@{}}ChatGPT\\ \end{tabular}} \\ \hline
        \textbf{1} & 4.88 & 0.84 \\
        \textbf{2} & 6.58 & 1.28 \\
        \textbf{3} & 5.97 & 1.29 \\
        \textbf{4} & 5.44 & 1.84 \\
        \textbf{5+} & 5.46 & 0.81 \\ \hline
        \textbf{Overall} & \textbf{5.77} & \textbf{1.18} \\ \hline
        \end{tabular}
    }}
    \captionof{table}{Exact match of baseline approach and ChatGPT. }
    \label{tab10}
    \end{minipage}
 \hfill
 \begin{minipage}{0.65\textwidth} 
    \centering
    \fontsize{9pt}{\baselineskip}\selectfont 
    {\color{black}\resizebox{\textwidth}{!}{
        \begin{tabular}{lllllll}
        \hline
        \multicolumn{1}{c}{\textbf{\begin{tabular}[c]{@{}c@{}}\# of\\ Entities\end{tabular}}} & \textbf{\begin{tabular}[c]{@{}l@{}}Bart \\ Base\end{tabular}} & \textbf{\begin{tabular}[c]{@{}l@{}}T5 \\ Base\end{tabular}} & \textbf{\begin{tabular}[c]{@{}l@{}}T5 \\ Large\end{tabular}} & \textbf{\begin{tabular}[c]{@{}l@{}}Flan T5 \\ Base\end{tabular}} & \textbf{\begin{tabular}[c]{@{}l@{}}Tk Inst\\ Base\end{tabular}} & \textbf{\begin{tabular}[c]{@{}l@{}}Tk Inst\\ w. Inst\end{tabular}}\\ \hline
        \textbf{1} & 17.46 & 15.82 & 17.03 & 16.05 & 16.02 & 17.35 \\
        \textbf{2} & 23.31 & 22.12 & 23.52 & 22.56 & 22.23 & 24.04 \\
        \textbf{3} & 21.40 & 19.29 & 21.42 & 19.97 & 19.66 & 22.20 \\
        \textbf{4} & 21.76 & 20.80 & 23.39 & 20.74 & 20.66 & 23.61 \\
        \textbf{5+} & 21.43 & 18.38 & 20.42 & 18.21 & 17.90 & 22.09 \\ \hline
        \textbf{Overall} & \textbf{20.88} & \textbf{19.31} & \textbf{20.92} & \textbf{19.64} & \textbf{19.44} & \textbf{21.45} \\ \hline
        \end{tabular}
    }}
    \captionof{table}{Exact match scores of supervised learning models. Tk Inst w. Inst denotes instruction tuning TkInstruct. }
    \label{tab11}
 \end{minipage}

\end{figure}

\begin{figure}[t!] 
\begin{minipage}{0.30\textwidth} 
    \centering
    \centering
    \fontsize{9pt}{\baselineskip}\selectfont 
    {\color{black}\resizebox{\textwidth}{!}{
        \begin{tabular}{lll}
        \hline
        \multicolumn{1}{c}{\textbf{\begin{tabular}[c]{@{}c@{}}\# of\\ Entities\end{tabular}}} & \textbf{Baseline} & \textbf{\begin{tabular}[c]{@{}l@{}}ChatGPT\\ \end{tabular}} \\ \hline
        \textbf{1} & 43.56 & 20.64 \\
        \textbf{2} & 45.40 & 17.76 \\
        \textbf{3} & 51.95 & 17.42 \\
        \textbf{4} & 55.94 & 18.54 \\
        \textbf{5+} & 64.26 & 24.99 \\ \hline
        \textbf{Overall} & \textbf{47.96} & \textbf{19.0} \\ \hline
        \end{tabular}
    }}
    \captionof{table}{No match of baseline approach and ChatGPT.}
    \label{tab12}
    \end{minipage}
 \hfill
 \begin{minipage}{0.65\textwidth} 
    \centering
    \fontsize{9pt}{\baselineskip}\selectfont 
    {\color{black}\resizebox{\textwidth}{!}{
        \begin{tabular}{lllllll}
        \hline
        \multicolumn{1}{c}{\textbf{\begin{tabular}[c]{@{}c@{}}\# of\\ Entities\end{tabular}}} & \textbf{\begin{tabular}[c]{@{}l@{}}Bart \\ Base\end{tabular}} & \textbf{\begin{tabular}[c]{@{}l@{}}T5 \\ Base\end{tabular}} & \textbf{\begin{tabular}[c]{@{}l@{}}T5 \\ Large\end{tabular}} & \textbf{\begin{tabular}[c]{@{}l@{}}Flan T5 \\ Base\end{tabular}} & \textbf{\begin{tabular}[c]{@{}l@{}}Tk Inst\\ Base\end{tabular}} & \textbf{\begin{tabular}[c]{@{}l@{}}Tk Inst\\ w. Inst\end{tabular}} \\ \hline
        \textbf{1} & 25.44 & 25.46 & 24.65 & 25.31 & 25.44 & 24.44 \\
        \textbf{2} & 23.39 & 23.68 & 22.73 & 23.60 & 23.62 & 22.91 \\
        \textbf{3} & 21.39 & 22.17 & 21.29 & 22.50 & 22.53 & 20.17 \\
        \textbf{4} & 23.18 & 23.15 & 21.94 & 23.40 & 23.05 & 21.51 \\
        \textbf{5+} & 24.88 & 26.74 & 27.07 & 29.51 & 28.99 & 23.83 \\ \hline
        \textbf{Overall} & \textbf{23.74} & \textbf{24.09} & \textbf{23.24} & \textbf{24.24} & \textbf{24.23} & \textbf{22.83} \\ \hline
        \end{tabular}
    }}
    \captionof{table}{No match scores of supervised learning models. Tk Inst w. Inst denotes instruction tuning TkInstruct.}
    \label{tab13}
 \end{minipage}

\end{figure}

\paragraph{No Match:} The baseline results are consistently worse than ChatGPT, as shown in Table \ref{tab12}. 
The no-match score for the baseline is more than twice as compared to ChatGPT (47.96\% vs. 19\%). 
As shown in Table \ref{tab13}, the no-match score for supervised learning models is also around 20\%. 
This could again be attributed to the recall scores, as all the supervised models and ChatGPT had recall scores of around 48.
Table \ref{tab14} gives a few examples of both exact and no matches by the Baseline method.

\begin{wraptable}{r}{0.6\textwidth}
\centering
\fontsize{9pt}{\baselineskip}\selectfont 
\renewcommand\tabcolsep{1pt} 
\renewcommand\arraystretch{1} 
\resizebox{0.6\textwidth}{!}
{
    \begin{tabular}{|l|l|l|}
    \hline
    \multicolumn{1}{|c|}{\textbf{Subject}} & \multicolumn{1}{c|}{\textbf{Relation}} & \multicolumn{1}{c|}{\textbf{Object}} \\ \hline
    \multicolumn{3}{|c|}{\textbf{Stanford OpenIE \cite{angeli2015leveraging}}}                                                                        \\ \hline
    Premium receivables                    & are reported                           & net of allowance                     \\ \hline
    fair value                             & received at                            & time of transactions                 \\ \hline
    we                                     & granted                                & approximately 0.3 stock units        \\ \hline
    variable                               & earn out                               & consideration                        \\ \hline
    \multicolumn{3}{|c|}{\textbf{Allen AI OpenIE \cite{Stanovsky2018SupervisedOI}  }}                                                                         \\ \hline
    Not Found                              & are                                    & Not Found                            \\ \hline
    the collateral                         & received at                            & at the time of the transactions      \\ \hline
    Not Found                              & restricted                             & stock units                          \\ \hline
    Not Found                              & estimated                              & value                                \\ \hline
    \end{tabular}
}
\caption{Responses of different OpenIE approaches on EDGAR10-Q examples in Table \ref{tab15}. }
\label{tab15}
\end{wraptable}

\section{Analysis}
In this section, compare our approach with traditional OpenIE approaches and highlight the differences between them. 
We also observe the effect of instruction tuning on the dataset and compare its performance. 
We finally explore the effects of the dataset with respect to different downstream tasks by using the models pre-finetuned on EDGAR on different downstream tasks.

\subsection{\name{} vs. OpenIE}
\label{open_ie_comparison}

\begin{table*}[t!]
\centering
\resizebox{\linewidth}{!}
{
    \begin{tabular}{lllll}
    \hline
     &
      \multicolumn{1}{c|}{\textbf{Sentence}} &
      \textbf{Entity} &
      \textbf{Labels} &
      \textbf{Baseline} \\ \hline
    \multicolumn{5}{c}{\textbf{Instances of Exact Match}} \\ \hline
    S1 &
      \multicolumn{1}{l|}{\begin{tabular}[c]{@{}l@{}}Premium receivables are reported net of an allowance for \\ doubtful accounts of \$250 and \$237 at September 30, 2020 \\ and December 31, 2019, respectively.\end{tabular}} &
      \begin{tabular}[c]{@{}l@{}}\$250 \\ and \\ \$237\end{tabular} &
      \begin{tabular}[c]{@{}l@{}}premium\\ receivable\end{tabular} &
      \begin{tabular}[c]{@{}l@{}}premium\\ receivable\end{tabular} \\ \hline
    S2 &
      \multicolumn{1}{l|}{\begin{tabular}[c]{@{}l@{}}The fair value of the collateral received at the time of the \\ transactions amounted to \$1,019 and \$351 at \\ September 30, 2020 and December 31, 2019, respectively.\end{tabular}} &
      \begin{tabular}[c]{@{}l@{}}\$1,019\\  and \\ \$351\end{tabular} &
      \begin{tabular}[c]{@{}l@{}}fair value\\ of \\ collateral\end{tabular} &
      \begin{tabular}[c]{@{}l@{}}fair value\\ of \\ collateral\end{tabular} \\ \hline
    \multicolumn{5}{c}{\textbf{Instances of No Match}} \\ \hline
    S3 &
      \multicolumn{1}{l|}{\begin{tabular}[c]{@{}l@{}}During the nine months ended September 30, 2020, we granted \\ approximately 0.3 restricted stock units that are contingent upon \\ us achieving earnings targets over the three year period from \\ 2020 to 2022\end{tabular}} &
      0.3 &
      \begin{tabular}[c]{@{}l@{}}grants\\ in\\ period\end{tabular} &
      \begin{tabular}[c]{@{}l@{}}restricted\\ stock\\ units\end{tabular} \\ \hline
    S4 &
      \multicolumn{1}{l|}{\begin{tabular}[c]{@{}l@{}}Certain selling equity holders elected to receive deferred, \\ variable earn out consideration with an estimated value \\ of \$21,500 over the rollover period of three years.\end{tabular}} &
      \$21,500 &
      \begin{tabular}[c]{@{}l@{}}earn\\ out\\ consideration\end{tabular} &
      \begin{tabular}[c]{@{}l@{}}estimated\\ value\end{tabular} \\ \hline
\end{tabular}
}
\caption{Instances of exact match and no match by the baseline approach. Column \textit{Baseline} denotes the responses generated by the baseline approach.}
\label{tab14}
\end{table*}


Traditionally, information extraction approaches from textual documents assume pre-specified relations for a given domain and employ crowd-sourcing or distant supervision approaches \cite{hoffmann2011knowledge,liu2016effective} to collect examples and train models for each type of relation. 
However, these approaches have a limitation in that they cannot extract unseen relations that were not observed or specified during training, rendering them impractical.
In contrast, Open Information Extraction (OpenIE) \cite{etzioni2008open} does not rely on pre-defined relations but extracts them on-the-fly as they are encountered. 
To compare our methods with existing OpenIE models, we evaluated Stanford's OpenIE and AllenNLP OpenIE models \cite{Stanovsky2018SupervisedOI} on a subset of the EDGAR10-Q dataset. 
Our findings indicate that Open IE models struggle when dealing with long-range dependencies.
We applied both OpenIE frameworks to the sentences shown in Table \ref{tab15} and present their results in Table \ref{tab14}. 
Notably, both frameworks failed to recognize any relations, contextual phrases, or entities.

\subsection{Marginal Improvement with Instruction Tuning} 
\begin{wrapfigure}{r}{0.5\textwidth}
\centering
\includegraphics[width= 0.5\textwidth, height= 4cm]{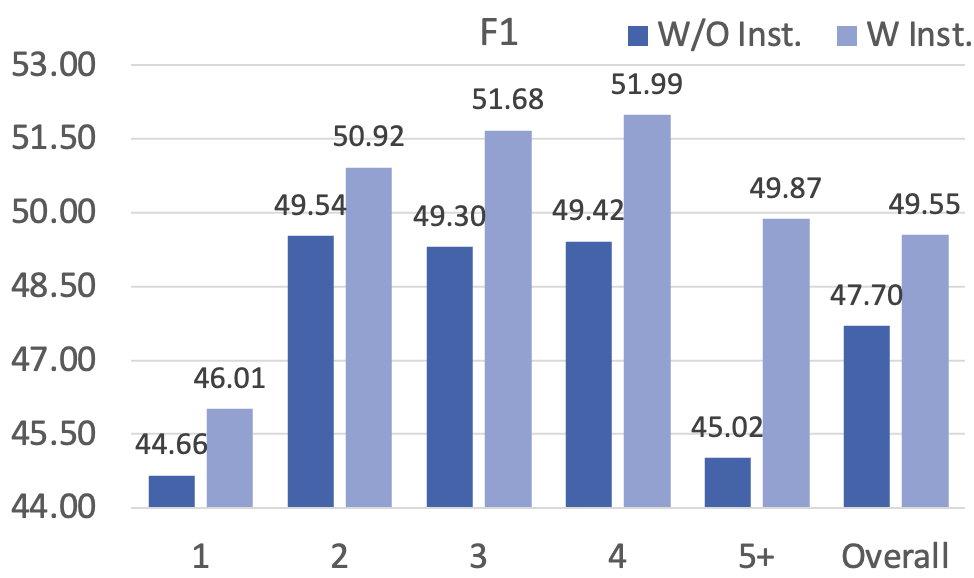}
\caption{ROUGE-L F1 scores for showing the effect of instruction tuning Tk Instruct vs conventional finetuning. X-axis denotes scores on different numbers of entities and overall score.}
\label{fig:instruction_tuning}
\end{wrapfigure}

Following works from instruction tuning \cite{wang-etal-2022-super}, we add instructions on the train data and instruction tune Tk-Instruct. 
Figure \ref{fig:instruction_tuning} showcases the performance increase across the entire test set. 
Across each sentence category, there is an increase of roughly 2\%, highlighting that instruction-tuned models with instruction data work well. 
The improvement is significant in sentences with 5+ categories where there is an increase of absolute 4\%. 

\subsection{Effect of EDGAR10-Q on downstream tasks}
\label{downstream_tasks}

To study the impact of EDGAR10-Q in the real world, we compare the effect of a model pre-finetuned on EDGAR10-Q vs a vanilla T5 model. 
We call T5 pre-finetuned on the dataset as EDGAR-T5.
Both EDGAR-T5 and vanilla T5 are then finetuned on three finance datasets; FiQA\cite{sinha2021impact}, FPB\cite{malo2014good}, and Headline \cite{10.1145/3184558.3192301} datasets. 
All the hyperparameters for vanilla T5 and Edgar-T5 were the same for a fair comparison\footnote{Hyperparameters are available in appendix}.
We use BloombergGPT-50B \cite{wu2023bloomberggpt} 10 shot score as the baseline for these tasks.
The splits used in downstream datasets and weighted F1 score were kept exactly the same as BloombergGPT for a fair comparison. 
EDGAR-T5 outperforms both vanilla T5 and BloombergGPT on all three downstream tasks and establishes SOTA results on all three of the tasks. 

\begin{wraptable}{r}{0.6\textwidth}
\centering
\fontsize{9pt}{\baselineskip}\selectfont 
\renewcommand\tabcolsep{1pt} 
\renewcommand\arraystretch{1} 
\resizebox{\linewidth}{!}
{
            \begin{tabular}{l|clclc}
            \hline
            \textbf{Dataset} & \textbf{Bloomberg GPT 50B} &  & \textbf{T5} &  & \textbf{EDGAR-T5} \\ \hline
            FiQA SA  & 75.07 &  & 74.89 &  & 80.42 \\
            FPB      & 51.07 &  & 55.77 &  & 79.69 \\
            Headline & 82.20 &  & 90.55 &  & 93.55 \\ \hline
            \end{tabular}
}
\caption{Comparison of EDGAR-T5 and Vanilla T5 on different finance related tasks.
Both models are 770M is size. 
BloombergGPT 50B is used as the baseline. 
Scores are weighted F1 as shown in BloombergGPT 50B.}
\label{tab16}
\end{wraptable}

In the FiQA sentiment analysis task, EDGAR-T5 outperforms vanilla T5 and BloombergGPT by 5.53 and 5.35\% points, respectively. 
In the FPB task, the performance gap increases significantly, and EDGAR-T5 outperforms vanilla and BloombergGPT by 23.92 and 28.62\% points, respectively. 
In the headline dataset, EDGAR-T5 got an F1 of 93.55\%, whereas vanilla T5 and BloombergGPT got F1 scores of 90.55\% and 82.20\%, respectively. 
The experiments suggest that the EDGAR10-Q dataset has led to an increase in the model's inherent ability for financial tasks.
We release all the finetuned models to the community for future use.

\section{Related Work}
\label{sec:related}

Several approaches have been developed for State-of-the-art NER detection \cite{chawla2021improving,li2019attention,luoma2020exploring,du2010using,zhu2018gram}.
Multiple approaches have been developed around various aspects of NER \cite{moon2018multimodal,amalvy2023role,li2020dice,kocaman2021biomedical,zhong2021frustratingly,Zhang2022OptimizingBF,shon2022slue,wang2022deepstruct}.
Etzioni et al. \cite{etzioni2008open} introduced a schemaless approach for extracting facts from text, focusing on relation extraction using OpenIE. 
However, this approach assumes relations between two entities, which poses challenges for financial data. 
Levy et al. \cite{levy2017zero} used a zero-shot approach to train MRC model on templatized questions and inferenced it on unseen relations.
Li et al. \cite{li2019entity} formalizes relation extraction as multi-turn question answering.  
Miwa et al. \cite{miwa2016end} jointly extracted entities and relations using neural networks, but performance suffers on unseen relations. 
Li and Ji \cite{li-ji-2014-incremental} use perceptron and efficient beam search to extract entities and relations.
Sun et al. \cite{sun-etal-2018-extracting} build on the previously mentioned framework and uses a joint learning framework and a flexible global loss function to capture the interactions of the entities and their relationships.
McCann et al. \cite{mccann2018natural} introduced decaNLP, addressing multiple challenges including relation extraction and question answering. 
Various frameworks like Stanford CoreNLP's NER, Spacy, NLTK, and Flair \cite{manning2014stanford,spacy,loper2002nltk,akbik2019flair} are available for entity extraction.
We aim to extract entities and their contexts, a more complex scenario than relation-based approaches.

\section{Conclusion}
\label{sec:conclusion}
In this paper, we introduced the Context-NER task, which aims to bridge the gap between existing NER tasks by extracting relevant phrases for entities. We also presented the EDGAR10-Q dataset, which is a large and complex finance dataset, providing a valuable resource for research in this domain. Through our baseline approach, we demonstrated the feasibility of solving the Context-NER task and conducted extensive experiments to evaluate our method's performance. Our comparison with GPT-3 showcased the challenges posed by the dataset. Additionally, we explored a supervised setting by finetuning pre-trained language models, which showed promising results. We believe that the introduction of the EDGAR10-Q dataset and our study will encourage further investigation and advancements in this field.

\section{Limitations and Future Work}
\label{limitations_and_future_work}
To advance our work, there are several promising directions for future research. 
Due to limited computational resources, we were unable to finetune large models (greater than 1B) on the EDGAR10-Q dataset. 
Elaborate experimentation could be conducted using other instruction-based or chain-of-thought reasoning on the EDGAR10-Q dataset. 
Future work should consider leveraging more powerful models to potentially achieve higher scores on this dataset. 
Furthermore, our evaluation set for ChatGPT was smaller than the actual test set due to budget constraints. 
Consequently, a few-shot evaluation of ChatGPT was omitted, investigating this avenue can lead to interesting results. 
Lastly, expanding the dataset to include reports from different markets and more recent years would enable researchers to explore the generalizability and temporal dynamics of the task.


\section{Ethical Considerations}

We have verified that all licenses of source documents used in this document allow their use, modification, and redistribution in a research context. 
There were no real-life names in the data set.  
No particular sociopolitical bias is emphasized or reduced specifically by our methods.

\section*{Acknowledgement}

The authors thank Arizona State University's Agave Research Computing cluster. 
The Cluster was used to create the dataset, run the baseline approach, and run the GPT-3 evaluation as well. 
Authors also extend their gratitude to Tarun Kumar, Amogh Badugu, Tamanna Agrawal and Dr. Himanshu Sharad Bhatt for their contributions.

\printbibliography

\section*{Checklist}


\begin{enumerate}

\item For all authors...
\begin{enumerate}
  \item Do the main claims made in the abstract and introduction accurately reflect the paper's contributions and scope? \answerYes{}
  
  \item Did you describe the limitations of your work?\answerYes{See \S \ref{sec:conclusion}}
  
  \item Did you discuss any potential negative societal impacts of your work?
\answerNo{We don’t expect negative societal impacts as a direct result of the contributions in our paper}
  \item Have you read the ethics review guidelines and ensured that your paper conforms to them? \answerYes{}
\end{enumerate}

\item If you are including theoretical results...
\begin{enumerate}
  \item Did you state the full set of assumptions of all theoretical results?\answerNA{}
	\item Did you include complete proofs of all theoretical results?\answerNA{}
\end{enumerate}

\item If you ran experiments (e.g. for benchmarks)...
\begin{enumerate}
  \item Did you include the code, data, and instructions needed to reproduce the main experimental results (either in the supplemental material or as a URL)?
    \answerYes{We included dataset, the script to generate it, finetuned model, and results in the supplementary material. ChatGPT results are reproducible as well.}
  \item Did you specify all the training details (e.g., data splits, hyperparameters, how they were chosen)?
    \answerYes{Please refer to \S \ref{experimental_setup}}
	\item Did you report error bars (e.g., with respect to the random seed after running experiments multiple times)?
    \answerNA{}
	\item Did you include the total amount of compute and the type of resources used (e.g., type of GPUs, internal cluster, or cloud provider)?
    \answerYes{Please refer to \S \ref{experimental_setup} for the type of resources, but we did not estimate the total amount of compute.}
\end{enumerate}

\item If you are using existing assets (e.g., code, data, models) or curating/releasing new assets...
\begin{enumerate}
  \item If your work uses existing assets, did you cite the creators?
    \answerYes{}
  \item Did you mention the license of the assets?
    \answerNo{The authors who created those artifacts already stated that in their paper. There was no additional input to add from our end. }
  \item Did you include any new assets either in the supplemental material or as a URL?
    \answerYes{We introduced new assets, and their respective GitHub and hugging face links were mentioned.}
  \item Did you discuss whether and how consent was obtained from people whose data you're using/curating?
    \answerNo{The authors of BART, T5, FlanT5, and Tk-Instruct had already released their models for public use along with their licenses. No other additional info was required from our end.}
  \item Did you discuss whether the data you are using/curating contains personally identifiable information or offensive content?
    \answerYes{The collected data does not contain personally
identifiable information or offensive content.}
\end{enumerate}

\item If you used crowdsourcing or conducted research with human subjects...
\begin{enumerate}
  \item Did you include the full text of instructions given to participants and screenshots, if applicable?
    \answerNA{}
  \item Did you describe any potential participant risks, with links to Institutional Review Board (IRB) approvals, if applicable?
    \answerNA{}
  \item Did you include the estimated hourly wage paid to participants and the total amount spent on participant compensation?
    \answerNA{}
\end{enumerate}

\end{enumerate}

\clearpage


\clearpage

\appendix
\section*{Appendix}

\section{Dataset Checklist}

\begin{enumerate}

\item \textbf{Dataset documentation and intended uses. Recommended documentation frameworks include datasheets for datasets, dataset nutrition labels, data statements for NLP, and accountability frameworks.} 
\answerYes{
We have provided a datasheet for the dataset (see \S\ref{datacard}). 
The intended use is to enable research on financial Natural Language Processing. 
Any usage for direct use or decision-making without review and supervision by professionals is out of scope. }

\item \textbf{URL to website/platform where the dataset/benchmark can be viewed and downloaded by the reviewers.} 
\answerYes{
The code required to scrape datasets, datasets, and machine learning experiments outlined in this manuscript is available on the GitHub code repository.
The  is also available at \url{https://huggingface.co/datasets/him1411/EDGAR10-Q}.\\
All the finetuned model weights are: available at \\ \url{https://huggingface.co/him1411/EDGAR-Tk-Inst-Large-Inst-Tune}\\
\url{https://huggingface.co/him1411/EDGAR-T5-Large}\\
\url{https://huggingface.co/him1411/EDGAR-Tk-Instruct-Large}\\
\url{https://huggingface.co/him1411/EDGAR-Flan-T5-Large} \\ 
\url{https://huggingface.co/him1411/EDGAR-BART-Base} \\  
\url{https://huggingface.co/him1411/EDGAR-T5-base}}

\item \textbf{Author statement that they bear all responsibility in case of violation of rights, etc., and confirmation of the data license.} 
\answerYes{see \S\ref{responsibility}. }

\item \textbf{Hosting, licensing, and maintenance plan. The choice of hosting platform is yours, as long as you ensure access to the data (possibly through a curated interface) and will provide the necessary maintenance.}
\answerYes{All code is hosted on GitHub at the repository linked above. 
We have released all dataset-related software under MIT License. 
EDGAR10-Q is an active open source project that is maintained by an international community of volunteers and the first two authors of the paper (Himanshu Gupta and Shreyas Verma).}

\item \textbf{Links to access the dataset and its metadata.} \answerYes{
see our project GitHub for dataset code, models, and metadata.}

\item \textbf{The dataset itself should ideally use an open and widely used data format. Provide a detailed explanation of how the dataset can be read. For simulation environments, use existing frameworks or explain how they can be used.} 
\answerYes{EDGAR10-Q is added to using Hugging Face's datasets hub to support easy integration into existing machine learning workflows. See \S\ref{schema} for details on the standardized schema to permit easier reuse. }

\item \textbf{Long-term preservation} \answerYes{The dataset is present in both Github and the hugging face dataset hub for its long-term preservation.}

\item \textbf{Explicit license} \answerYes{All code for is released under MIT License. All dataset licensing remains the same as the source. See \S\ref{datacard} for complete licensing information for all datasets in ours.}

\end{enumerate}

\section{Dataset Documentation}

\subsection{Dataset documentation and intended uses}

\begin{enumerate}
 \item \textbf{For what purpose was the dataset created? Was there a specific task in mind? Was there a specific gap that needed to be filled? Please provide a description.} 
 The EDGAR10-Q dataset was created to address the task of finding relevant phrases associated with entities, which is formally defined as \name{}. 
 The dataset was created to enable a systematic study of this task and to provide a challenging benchmark for state-of-the-art models. 
 The task involves generating a relevant phrase that describes an entity in a sentence, which may or may not be present in the sentence. 
 The dataset consists of quarterly and annual financial reports of publicly traded LLCs and contains complex sentences that are significantly longer than those found in typical benchmark datasets. 
 The dataset also contains several numerical entities, making it particularly challenging for models to retrieve the relevant phrases without adequate domain knowledge. 
 The creation of this dataset fills a gap in the research by providing a quality dataset for the \name{} task, enabling researchers to explore this important problem in greater depth.

 \item \textbf{Who created this dataset (e.g., which team, research group) and on behalf of which entity (e.g., 15 company, institution, organization)?} 
 The first two authors (Himanshu Gupta and Shreyas Verma were responsible for the creation of this dataset. 
 This was done as an independent study project by the two authors.

\item \textbf{Who funded the creation of the dataset?} \answerNA{}
\item \textbf{Any other comments?} \answerNA{}
  
\end{enumerate}

\subsection{Composition}

\begin{enumerate}

 \item \textbf{What do the instances that comprise the dataset represent (e.g., documents, photos, people, countries)? Are there multiple types of instances (e.g., movies, users, and ratings; people and interactions between them; nodes and edges)? Please provide a description.} 
 Each row represents a sentence that was present in the quarterly or annual financial report of an organization. See \S \ref{schema} for more details. 
  
  \item \textbf{How many instances are there in total (of each type, if appropriate)? Does the dataset contain all possible instances or is it a sample (not necessarily random) of instances from a larger set? If the dataset is a sample, then what is the larger set? Is the sample representative of the larger set (e.g., geographic coverage)? If so, please describe how this representativeness was validated/verified. If it is not representative of the larger set, please describe why not (e.g., to cover a more diverse range of instances, because instances were withheld or unavailable).}   
  EDGAR10-Q consists of approximately 1M instances that have 2.78M entities. 
  All instances from the dataset were used, and no sample was left out. 
  Irrelevant sentences were removed while cleaning the dataset. 
  As each sentence had multiple entities, during supervised training, each row had the sentence repeat multiple times with different entities present in the sentence and the corresponding phrase describing the entity. 
  The train split consists of 1.49M instances, and both the eval and test set consists of 187K instances each.
  
  \item \textbf{What data does each instance consist of? “Raw” data (e.g., unprocessed text or images) or features? In either case, please provide a description.} 
  This is described in \S \ref{schema}.

\item \textbf{Is there a label or target associated with each instance?If so, please provide a description.} 
This is described in \S \ref{sec:dataset} and \S \ref{schema}.

\item \textbf{Is any information missing from individual instances? If so, please provide a description, explaining why this information is missing (e.g., because it was unavailable). This does not include intentionally removed information, but might include, e.g., redacted text.}  
The meta dataset released has all the information (see \S \ref{schema}), but the supervised learning task uses only the entity, the sentence as input, and the phrase describing the entity as the output.
Moreover, any PII (personally identifiable information) has been removed from the dataset. 

\item \textbf{Are relationships between individual instances made explicit (e.g., users’ movie ratings, social network links)? If so, please describe how these relationships are made explicit.} \answerNA{}

\item \textbf{Are there recommended data splits (e.g., training, development/validation,testing)? If so, please provide a description of these splits, explaining the rationale behind them.} 
The training, test, and validation split have been standardized and will be shared with the community along with the dataset for reproducing our results. 
We randomly split the data into these splits in the ratio of 80\% 10\% and 10\%.

\item \textbf{Are there any errors, sources of noise, or redundancies in the dataset? If so, please provide a description.}
Since the dataset has been sourced from annual and quarterly reports, we observe spelling mistakes, grammatical mistakes, etc. 

\item \textbf{Is the dataset self-contained, or does it link to or otherwise rely on external resources (e.g., websites, tweets, other datasets)?} 
EDGAR10-Q is self-contained.

\item \textbf{Does the dataset contain data that might be considered confidential (e.g., data that is protected by legal privilege or by doctor-patient confidentiality, data that includes the content of individuals’ non-public communications)?If so, please provide a description.} 
No. We completely anonymized the dataset to protect the privacy of the users. 

\item \textbf{Does the dataset contain data that, if viewed directly, might be offensive, insulting, threatening, or might otherwise cause anxiety? If so, please describe why.} No.

\item \textbf{Does the dataset relate to people? If not, you may skip the remaining questions in this section.} No. In particular, the dataset does contain any Personally Identifiable Information that could be mapped back to an individual.
  
\end{enumerate}

\subsection{Collection process}
\begin{enumerate}

\item \textbf{How was the data associated with each instance acquired? Was the data directly observable (e.g., raw text, movie ratings), reported by subjects (e.g., survey responses), or indirectly inferred/derived from other data (e.g., part-of-speech tags, model-based guesses for age or language)? If data was reported by subjects or indirectly inferred/derived from other data,
was the data validated/verified? If so, please describe how.} 
See \S \ref{creation} for details

\item \textbf{What mechanisms or procedures were used to collect the data (e.g., hardware apparatus or sensor, manual human curation, software program, software API)? How were these mechanisms or procedures validated?} 
Since the dataset was created by scraping documents, no special apparatus was required. 
The data creation is described in \S \ref{creation}.

\item \textbf{If the dataset is a sample from a larger set, what was the sampling strategy (e.g., deterministic, probabilistic with specific sampling probabilities)?} \answerNA{}

\item \textbf{Over what timeframe was the data collected? Does this timeframe match the creation timeframe of the data associated with the instances (e.g., recent crawl of old news articles)? If not, please describe the time-frame in which the data associated with the instances was created.} 
The dataset contains data from 2 financial years and required 3 days to scrape and clean. This time also included baseline creation as well.

\item \textbf{Were any ethical review processes conducted (e.g., by an institutional review board)? If so, please provide a description of these review processes, including the outcomes, as well as a link or other access point to any supporting documentation.} \answerNA{}

\item \textbf{Does the dataset relate to people? If not, you may skip the remainder of the questions in this
section.} \answerNA{}

\end{enumerate}

\subsection{Preprocessing/cleaning/labeling}

\begin{enumerate}
\item \textbf{Was any preprocessing/cleaning/labeling of the data done (e.g., discretization or bucketing, tokenization, part-of-speech tagging, SIFT feature extraction, removal of instances, processing of missing values)? If so, please provide a description. If not, you may skip the remainder of the
questions in this section.}  Please refer to \S \ref{creation} for this. 

\item \textbf{Was the “raw” data saved in addition to the preprocessed/cleaned/labeled data (e.g., to support
unanticipated future uses)? If so, please provide a link or other access point to the “raw” data.}
Raw data was not saved and cleaned on the fly to prevent misuse. 

\item \textbf{Is the software used to preprocess/clean/label the instances available? 
If so, please provide a link or other access point.} The code was written in Python and is present in the Github shared. 

\end{enumerate}

\subsection{Uses}

\begin{enumerate}
\item \textbf{Has the dataset been used for any tasks already? If so, please provide a description} Since we are proposing a new task and the dataset along with it, to the best of our knowledge, no other tasks are currently using it.

\item \textbf{Is there a repository that links to any or all papers or systems that use the dataset? If so, please provide a link or other access point.} \answerNA{}

\item \textbf{What (other) tasks could the dataset be used for?} The dataset could be used for traditional NER and its related tasks. 

\item \textbf{Is there anything about the composition of the dataset or the way it was collected and preprocessed/cleaned/labeled that might impact future uses? For example, is there anything that a future user might need to know to avoid uses that could result in unfair treatment of individuals or groups (e.g., stereotyping, quality of service issues) or other undesirable harms (e.g., financial harms, legal risks) If so, please provide a description. Is there anything a future user could do to mitigate these undesirable harms?} Dataset captures only the US market. Future uses of the dataset could be increased by including data from other markets. 

\item \textbf{Are there tasks for which the dataset should not be used? If so, please provide a description.} \answerNA{}

\end{enumerate}

\subsection{Distribution}

\begin{enumerate}

\item \textbf{Will the dataset be distributed to third parties outside of the entity (e.g., company, institution, organization) on behalf of which the dataset was created? If so, please provide a description.} The dataset and the codebase are available on GitHub. 

\item \textbf{How will the dataset will be distributed (e.g., tarball on website, API, GitHub)? Does the dataset have a digital object identifier (DOI)?} The dataset is available on the GitHub repository and huggingface. 

\item \textbf{When will the dataset be distributed?} The dataset is already publically available on both GitHub and huggingface. 

\item \textbf{Will the dataset be distributed under a copyright or other intellectual property (IP) license, and/or under applicable terms of use (ToU)? If so, please describe this license and/or ToU, and provide a link or other access point to, or otherwise reproduce, any relevant licensing terms or ToU, as well as any fees associated with these restrictions.} 
The dataset and the codebase are under the MIT license.

\item \textbf{Have any third parties imposed IP-based or other restrictions on the data associated with the instances? If so, please describe these restrictions, and provide a link or other access point to, or otherwise reproduce, any relevant licensing terms, as well as any fees associated with these restrictions.} No.

\item \textbf{Do any export controls or other regulatory restrictions apply to the dataset or to individual instances? If so, please describe these restrictions, and provide a link or other access point to, or otherwise reproduce, any supporting documentation.} No. 

\end{enumerate}

\subsection{Maintenance}
\begin{enumerate}

\item \textbf{Who is supporting/hosting/maintaining the dataset?} Authors of this work bear all responsibility in case of violation of rights. 

\item \textbf{How can the owner/curator/manager of the dataset be contacted (e.g., email address)?} If you wish to extend or contribute to our dataset, please contact us via email.

\item \textbf{Is there an erratum? If so, please provide a link or other access point.} Any updates to the dataset wiil be shared via GitHub. 

\item \textbf{Will the dataset be updated (e.g., to correct labeling errors, add new instances, delete instances)? If so, please describe how often, by whom, and how updates will be communicated to users (e.g., mailing list,GitHub)?} If we find inconsistencies in the dataset or extend the dataset, we will release the new version via GitHub and Huggingface. 

\item \textbf{If the dataset relates to people, are there applicable limits on the retention of the data associated with the instances (e.g., were individuals in question told that their data would be retained for a fixed period of time and then deleted)?} \answerNA{}

\item \textbf{Will older versions of the dataset continue to be supported/hosted/maintained? If so, please describe how. If not, please describe how its obsolescence will be communicated to users} All versions will continue to be supported and maintained on GitHub and HuggingFace. We will post the updates on the GitHub repository. 

\item  \textbf{if others want to extend/augment/build on/contribute to the dataset, is there a mechanism for them to do so? If so, please provide a description. Will these contributions be validated/verified? If so, please describe how. If not, why not? Is there a process for communicating/distributing these contributions to other users? If so, please provide a description.} Yes. Please contact the authors of this paper to build upon this dataset 
\end{enumerate}

\subsection{Responsibility}
\label{responsibility}
The Authors bear all responsibility in case of violation of rights, caused by the development and release of ours. 
The code is released under MIT License. 
We confirm that the dataset is licensed under a Creative Commons License Family.

\section{Extended Related Work}
\label{sec:extended_related}

LMs and Deep learning methods have been used for plethora of downstream tasks for ling time \cite{Yin2018LearningTR,li2017code,Das2015ContextualCC,10.1007/978-3-030-45778-5_27, 10.1007/978-3-030-75075-6_10,10.1007/978-3-030-69143-1_18,10.1007/978-3-030-87013-3_4,Husain2019CodeSearchNetCE,feng-etal-2020-codebert,Vijayakumar2018NeuralGuidedDS}.
Several recent works have leveraged NLP methods and simple sampling methods for different downstream results \cite{10.1145/3287075,Alon2018code2vecLD,allamanis2017learning,balog2016deepcoder,10.1007/978-981-16-8892-8_29,10.1007/978-981-16-8892-8_46,8877082}.
The study of whether existing LMs can understand instructions by \cite{Efrat2020TheTT} has motivated a range of subsequent works. 
For instance, \cite{hase-bansal-2022-models}, \cite{ye-ren-2021-learning}, and \cite{zhong-etal-2021-adapting-language} have proposed different methods to demonstrate that language models can follow instructions. 
\cite{weller-etal-2020-learning} developed a framework that focuses on developing NLP systems that solve new tasks after reading their descriptions. 
\cite{mishra2021cross} proposed natural language instructions for cross-task generalization of LMs. 
PromptSource and FLAN \cite{weifinetuned, sanh2021multitask} were built to leverage instructions and achieve zero-shot generalization on unseen tasks. 
Moreover, \cite{parmar-etal-2022-boxbart} shows the effectiveness of instructions in multi-task settings for the biomedical domain. 
\cite{mishra-etal-2022-reframing} discussed the impact of task instruction reframing on model response, while \cite{min-etal-2022-metaicl} introduced a framework to better understand in-context learning. 
Additionally, \cite{ouyang2022training} proposed the InstructGPT model, which is fine-tuned with human feedback to follow instructions. 
\cite{gupta2022john} showed that adding knowledge with instruction helps LMs understand the context better.
\cite{Wang2022InstructionNERAM} developed an instruction-based multi-task framework for few-shot Named Entity Recognition (NER) tasks. 
Furthermore, several approaches have been proposed to improve model performance using instructions, including \cite{Wu2021AICT, liu2022few, luo2022biotabqa, kuznia-etal-2022-less, patel-etal-2022-question, mishra2022help,puri-etal-2023-many}.

\section{EDGAR10-Q Datacard}
\label{datacard}

\textbf{Dataset Description:} EDGAR10-Q is a dataset for introducing the ContextNER task, which aims to generate the relevant context of entities in a sentence. 
The generated context can be phrases describing the entity but not necessarily present in the sentence. 
EDGAR10-Q dataset is created from annual and quarterly reports of the top 1500 publicly traded companies. 
It is the largest dataset in terms of the number of sentences (1M), entities (2.8M), and the average number of tokens per sentence (35).
The train set has a total $1,499,079$ instances. Eval and Test Set have $187,383$ instances each.

\textbf{Homepage:} \url{https://github.com/him1411/edgar10q-dataset}

\textbf{URL:} \url{https://huggingface.co/datasets/him1411/EDGAR10-Q}

\textbf{Licensing:} MIT License Family

\textbf{Languages:} English

\textbf{Tasks:} NER, Context NER

\textbf{Splits:} train, validation, test

\section{Dataset Creation}
\label{creation}
The URL \url{https://github.com/him1411/edgar10q-dataset/blob/main/dataset_generation_and_baseline.py} contains the data extraction and baseline generation code. The process to extract data is described below: 

The code starts from the function called driver\_writer\_func, which takes five arguments: 
company\_name, cik, start\_date, base\_folder, and dest\_folder. 
The function performs the following steps:

\begin{enumerate}
    \item Deletes the base\_folder directory and creates a new one.
    \item Creates a dest\_folder directory if it does not already exist.
    \item Calls a function get\_all\_submissions with arguments cik, start\_date, base\_folder, and company\_name to retrieve financial documents for the company.
    \item Parses the documents using IE\_Parser and structures the resulting data in a tabular format.
    \item Filters the data based on certain criteria such as text length and data type.
    \item Parses the text data to extract entities, phrases, and questions and answers using various functions such as sent\_parse, sentence\_entity\_flair, phrase\_extraction, and qa\_model (They are part of baseline extraction method and are explained in \S ).
    \item Writes the resulting data to a CSV file in the dest\_folder directory with the name company\_name.csv.
\end{enumerate}

\paragraph{get\_all\_submissions:} takes four arguments - cik (an integer), start\_date, base\_folder (a string), and company\_name (also a string). 
It first checks if the company\_name exists in a file called done\_comps, and if so, prints a message saying all files of the company have already been downloaded and returns None.
Next, it reads the contents of a file called DONE\_LIST if it exists, and assigns it to the variable done\_subs.
Then, it converts cik to a string data type, calls the function get\_accession\_numbers with cik, '10-K', and start\_date as arguments, and assigns the result to the variable subs\_10k. 
Similarly, it calls get\_accession\_numbers with cik, '10-Q', and start\_date as arguments and assigns the result to the variable subs\_10q. 
It then concatenates these two lists (subs\_10k and subs\_10q) into a new list called subms.
The function then logs the number of submissions made after start\_date by cik.
For each submission in the subms list, the function extracts the name and url of the JSON file associated with it. 
It then loads the JSON data into a dictionary called subm\_json using the json.loads() method. 
From this dictionary, it extracts the list of files associated with the submission and filters out those with a .txt file extension. 
It then selects the first .txt file and extracts its file name and url.
Next, it calls the function get\_meta\_data with the contents of the text file as an argument to extract metadata from the file. 
If successful, it assigns the submission type based on the extracted metadata. 
If the submission type folder doesn't exist in the base\_folder, it creates the necessary directory structure.
It then writes the contents of the text file to a file in the appropriate directory in the base\_folder, and appends the name of the submission to a file called DONE\_LIST. 
Finally, it appends the company\_name to a file called DONE\_COMP.

\paragraph{get\_accession\_numbers:} accepts three parameters: cik (a string), type (a string) and start\_date (a datetime object). 
It returns a list of accession numbers for a company with the specified Central Index Key (CIK) that have been filed with the Securities and Exchange Commission (SEC) after a specified date and of a specified type.
The function starts by constructing a URL based on the parameters passed in. 
The URL is used to fetch an HTML page containing a table of filing information. 
The function then processes the HTML page using BeautifulSoup to extract the relevant table, convert it to a pandas DataFrame, and filter the rows to those with filing dates greater than the specified start\_date. 
It then extracts the accession numbers from the filtered table, cleaning them up and returning them as a list.

\paragraph{get\_meta\_data:} accepts a string subm\_details\_text that contains the content of a submission details text file in the EDGAR database. 
It returns a dictionary containing the metadata for the submission.
The function starts by initializing an empty dictionary called meta\_data and two lists called running\_titles and running\_indents. 
It then loops over the rows of the subm\_details\_text string, splitting each row into segments using the colon (:) as a separator. 
If a row contains a single segment, the function assumes that it is a heading and adds it to running\_titles along with its indentation level, which is calculated by counting the number of tabs in the row. 
If a row contains two segments, the function assumes that it is a key-value pair and adds it to meta\_data using the deep\_set function to create nested dictionaries for the various levels of headings. 
The deep\_set function sets the value of a nested dictionary by walking the dictionary hierarchy according to the list of headings passed in and creating new dictionaries as needed. 
Finally, the function returns the meta\_data dictionary.

\section{Dataset Schema}
\label{schema}

\subsection{Raw Data}

\begin{figure*}[h!]
	\centering
	\includegraphics[width=\linewidth, height= 7 cm]{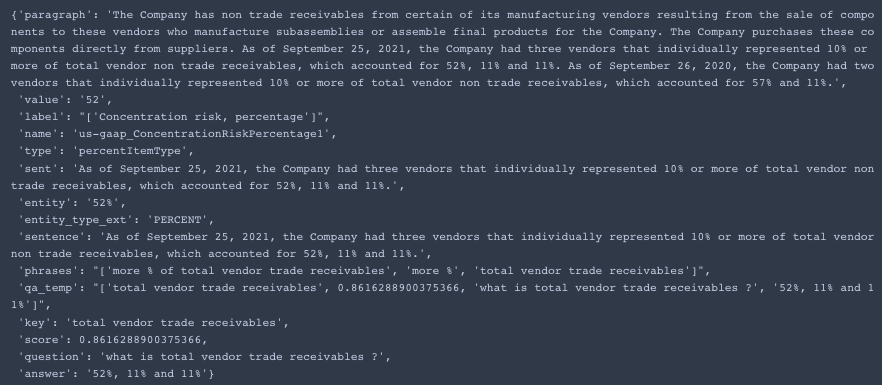}
	\caption{Illustration of the raw data in json format obtained after dataset collection.}
	\label{row_instance}
\end{figure*}

Figure \ref{row_instance} shows one instance of the raw dataset. The complete dataset is present in the GitHub repository. 
Each column is described below:

\begin{enumerate}
    \item paragraph: It contains the input string and the sentences surrounding it.
    \item value: The numerical value of the entity whose context is going to be extracted from the sentence 
    \item label: A list of phrases, which describe the entity. In this case, the phrases are: 'Concentration risk, percentage'.
    \item name: A string representing the name of the value, in this case, is 'us-gaap\_ConcentrationRiskPercentage1'.
    \item type: Description of the data type of the value, in this case, is 'percentItemType'.
    \item sent: The sentence that contains the entity.
    \item entity: Entity extracted using NER library'52\%'.
    \item entity\_type\_ext: The data type of the entity extracted using the NER library, which is 'PERCENT'.
    \item sentence: Cleaned version of sent.
    \item phrases: Phrases extracted from the phrase generation algorithm, including 'more \% of total vendor trade receivables', 'more \%', and 'total vendor trade receivables'.
    \item qa\_temp: List of questions that are formed using the phrases.
    \item key: The phrase whose question gave the correct answer. 
    \item score: The confidence score given by the answer, that is  0.8616288900375366 in this case.
    \item question: the question that gave the correct answer, that is, 'What is total vendor trade receivables ?' in this case.
    \item answer: String that represents the MRC output of the BERT model used in the baseline approach.
\end{enumerate}

\subsection{Supervised modeling data}

Supervised modeling data consisted of the entity concatenated with the sentence. The output is one of phrases from the labels. In this case, the input is: 
\textit{52\%.As of September 25, 2021, the Company had three vendors that individually represented 108 or more of total vendor non trade receivables, which accounted for 52\%, 11\%, and 11\%}.
The output for this sentence is:
\textit{Concentration risk}

\textbf{Supervised Training setup:} We finetune generative models (T5 Base, T5 Large \cite{2020t5}, BART Base \cite{lewis2020bart}, Flan-T5 Large \cite{Chung2022ScalingIL}, Tk-Instruct Large \cite{wang-etal-2022-super}) on the train split of the dataset. 

\textbf{Hyper parameters}: Train Batch Size: 8, Gradient Accumulation Steps: 8, Max Source Length: 512, Max Target Length: 128, Number of Epochs: 2, Warmup Steps: 100, Learning Rate: $\{5\}e{-}5$

\begin{algorithm*}[t!]
\SetStartEndCondition{ }{}{}%
\SetKwProg{Fn}{def}{\string:}{}
\SetKwFunction{Range}{range}
\SetKw{KwTo}{in}\SetKwFor{For}{for}{\string:}{}%
\SetKwIF{If}{ElseIf}{Else}{if}{:}{elif}{else:}{}%
\AlgoDontDisplayBlockMarkers
\SetAlgoNoEnd
\SetAlgoNoLine%
\SetKwInput{KwInput}{Input}                
\SetKwInput{KwOutput}{Output}              
\DontPrintSemicolon
\KwInput{Sentence}
\KwOutput{List of Phrases}

\SetKwFunction{FMain}{complex\_noun\_phrase\_extractor}
\SetKwFunction{FSum}{simple\_noun\_phrase\_extractor}

\SetKwProg{Fn}{Function}{:}{\KwRet}
\Fn
{\FSum{Sentence}}
  {
        doc = sequence\_of\_token(Sentence),phrase\_list = []    \;
        
        \For{token in Doc }
        {
            phrase = \textquotesingle \hspace{.25cm}  \textquotesingle  \;

            \If{token.head.pos in [Noun, Pronoun] and token.dep in [Object, Subject]  }
            {
                \For{subtoken in token.children }
                {
                    \lIf{subtoken.pos is Adj or subtoken.dep is Comp }
                    {
                        phrase += subtoken.text + \textquotesingle \hspace{.25cm} \textquotesingle
                    }
                }
                \lIf{len(phrase) is not 0}
                {
                    phrase += token.text
                }
            }
                
            \lIf{len(phrase) is not 0 and phrase doesnot have entities}
            {
                phrase\_list.append(phrase)
            }
        }
        \KwRet phrase\_list\;
  }
    
  \SetKwProg{Fn}{Function}{:}{}
  \Fn
  {\FMain{Sentence}}
  {
        doc = sequence\_of\_token(Sentence)   \;
        phrase\_list = []  \;
        
        \For{token in Doc }
        {
            \If{token.pos is Preposition}{phrase = \textquotesingle \hspace{.25cm}\textquotesingle

                \If{token.head.pos in [Noun, Pronoun]}
                {
                    \For{subtoken in token.head.children}
                    {
                        \If{subtoken.pos is Adj or subtoken.dep is Comp }
                        {
                            phrase += subtoken.text + \textquotesingle  \hspace{.15cm}  \textquotesingle \;
                        }
                    }
                    phrase += token.head.text + \textquotesingle  \hspace{.25cm}  \textquotesingle + token.text \;

                    \For{right\_tok in token.rights }
                    {
                        \If{right\_tok in [Noun, Pronoun]}
                        {
                          \For{subtoken in right\_tok.children }
                            {
                                \If{subtoken.pos is Adj or subtoken.dep is Comp }
                                {phrase += subtoken.text + \textquotesingle  \hspace{.25cm}  \textquotesingle 
                                }
                            }
                            phrase += \textquotesingle  \hspace{.25cm}  \textquotesingle + right\_tok.text  \;
                        }
                    }
                    \lIf{len(phrase) is $>$ 1 and phrase doesnot have entities}
                    {
                        phrase\_list.append(phrase)
                    }
                }
            }
        }
        \KwRet  phrase\_list \;
  } 
  \caption{Phrase Generation \textit{Pseudocode} }
  \label{algo1}
\end{algorithm*}

\section{Baseline approach }
\label{baseline}

\begin{figure*}
\centering
\includegraphics[width= 6cm, height= 9 cm]{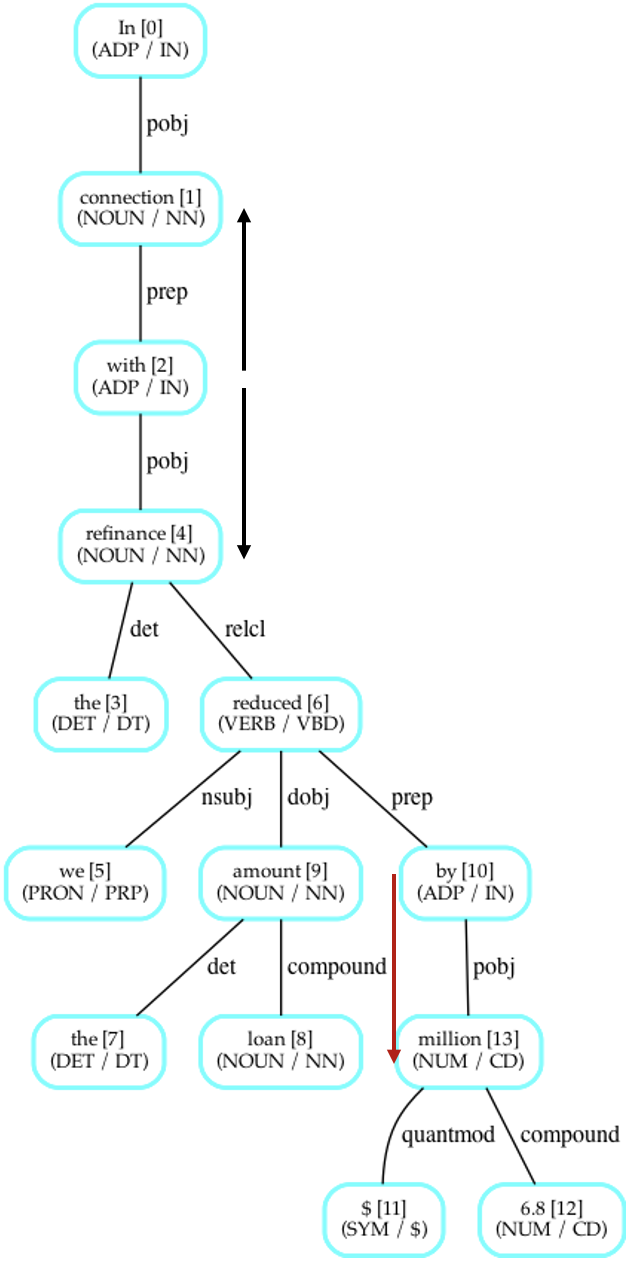}
\caption{Syntactic tree structure for extraction of simple and complex noun phrases. }
\label{syntactc_tree}
\end{figure*}

Consider the sentences: 

\begin{itemize}
    \item In October 2019, the Company increased the borrowing capacity on the revolving credit loan by \$33,000 increasing the available credit facility from \$60,000 to \$93,000.
    \item If the loan is paid during months 13-24 or 25-36 and then a penalty of 2\% and 1\%, respectively, of the loan balance will be charged on the date of repayment.
    \item The weighted-average remaining lease term and discount rate related to the Company’s lease liabilities as of September 26, 2020 were 10.3 years and 2.0\%, respectively. 
\end{itemize}

\subsection{Phrase Generation}

This paper presents a simple, yet efficient technique to extract entities and their descriptions from sentences. As shown in Figure \ref{fig:block}, it starts with data cleaning and entity extractions. A noun phrase (NP) \cite{stuart2013importance} includes a noun, a person, place, or thing, and the modifier that distinguishes it. Open IE is predicated on the idea that the relation (which is action verbs in most cases) is the central element of the study, from which all other considerations flow. However, in many cases, the verb is not helpful, particularly in financial data. Consider the sentence: "Deferred revenue for 2020 is \$20 billion." Like most financial records are of the form "is, was, be," etc., the verb "is" in this sentence is an auxiliary verb and does not describe any particular event or give any information about the entity. 

We extract two types of phrases from the sentences, namely simple and complex.  In simple phrase extraction, each sentence comprises subject-object and verb connecting them where Subject or Object is usually a noun or pronoun. After searching for a noun and pronoun, we check for any noun compound or adjective. On the other hand, for complex phrase extraction we first start with preposition extraction. We then follow similar steps as in simple phrase extraction to look for phrases in both left and right of the preposition. It has to be noted that simple phrases are not always found on both sides of the proposition. Algorithm \ref{algo1} further summarizes the process of simple and complex phrase extraction from the sentences.

Now we demonstrate the extraction of simple and complex noun phrases for the sentence, \textit{'In connection with the refinance we reduced the loan amount by \$6.8 million.'}. 
The syntactic tree for the above sentence is shown in Figure \ref{syntactc_tree}.
We search if the token's POS tag is a noun or pronoun as we are looking just for noun phrases. We also ensure that phrase lies either in the Subject or Object of the sentence to ensure we are skipping the relations. In this case, we got \textit{"amount"}  the first word of the phrase. After that, we iterate the node to see its children named subtoken in Algorithm \ref{algo1}. We search for subtoken's dependency relation with the token as a compound relation, or we search if the subtoken is an adjective. The intuition behind this is that if the subtoken and token have a compound relationship, they form a meaningful noun phrase. In this case, "amount" has a compound relationship with its subtoken \textit{"loan"} so they together form \textit{"loan amount"} as the meaningful noun phrase. Similar logic is followed for searching adjectives. Complex NPs are identified as series of noun phrases with a preposition separating them, so we start by identifying them. In this example, the preposition identified was \textit{"in"}. Then we iterate both up and down the node to find noun phrases that follow the same method mentioned above. The noun phrases identified from the top were \textit{"connection"} and the bottom was \textit{"refinance"}. The entire complex NP was formed as NP from top + preposition in the middle and + NP from below. The resultant was \textit{"connection with refinance"}.


\subsection{Machine Reading Comprehension Model}
\label{mrc_model}

This paper presents a zero-shot technique as we leverage the phrase generation to generate meaningful questions without further training of the machine reading comprehension (MRC) model. This allows our technique to be domain agnostic and thus can be easily expanded to newer domains. The process to leverage noun phrases to generate the questions and further using the MRC model to associate entities with their corresponding descriptions is described below: 

\begin{itemize}
    \item Each paragraph in the document is broken down into sentences. For each sentence, the following are extracted: Phrases (using simple and complex noun phrases described in Algorithm \ref{algo1}) and Entities using the Flair NER Model.
    \item On the basis of the entity type and the noun phrases, the questions are framed accordingly. For instance, if the entity found out was of type date, then the question would be "when is" + NP?. In our example, the question for the first sentence for \S \ref{baseline} would be "how much is borrowing capacity on revolving credit loan ?". 
    \item In instances where the entity type is of integer, float, or percent where appending "when is" or "how much is" does not give an advantage. For such cases, to keep the question generic we append "what is" to the noun phrase.  For example, the question for second sentence for \S \ref{baseline} is, "What is the loan balance?" was created based on the entity type of 2\% and 1\%.
    \item Once these questions are generated, they are fed into the MRC Model, and its answer is checked if it contains the entity. To give an example, in the 1st sentence, the following questions are created, and the model returns their corresponding answers and their confidence values:
    \begin{itemize}
        \item "How much is borrowing capacity on revolving credit loan?"  answer: "\$33,000", confidence score: 0.946
        \item "How much is borrowing capacity ?" answer: "\$33,000", confidence score: 0.824
        \item "How much is revolving credit loan ?" answer: "\$33,000", confidence score: 0.856
        \item "How much is available credit facility ?" answer: "\$60,000 to \$93,000", confidence score: 0.5762
    \end{itemize}
    If there are multiple questions whose answer has the entity, we select the question whose answer is of the highest confidence value. In the above example, "borrowing capacity on revolving credit loan" is chosen as the key for \$30,000, and "revolving credit loan" is chosen as the key for both \$60,000 to \$93,000.

    \item If the entity is not present in the response of the MRC model, the question is discarded. In the 2nd Sentence of Table 1, the following questions are created : 
    \begin{itemize}
        \item "What is penalty of \% ?"
        \item "What is loan balance ?"
    \end{itemize}
    None of them are returning "13-24 or 25-36", so the phrases "penalty of \%" and "loan balance" are discarded.
    
    \item There are instances where none of the generated questions returned an answer with the target entity or returned responses with a different entity as shown above. For those cases, we create the question "what is" entity?.  Here, its response would be considered as the key (opposite to the case above). In the 2nd sentence of the Table, none of the questions returned relevant answers, So the following questions were created: 
    \begin{itemize}
        \item "What is 13-24 or 25-36 ?"
        \item "What is 2\% and 1\% ?"
    \end{itemize}
    
    \item In the above cases, where questions are formed based on entities, the answers are checked if they have given any other entity as the answer.  For instance, the questions, "what is 2\% and 1\% ?" return "2" as the answer to the second sentence. If the cases mentioned above hold, then the response is discarded. Here all the cases to identify the noun phrase associated with the entity fail, so no answer is returned.
    
    \item If they do not fail, then the response is also considered a viable answer. For instance, In the 2nd sentence, the question was framed:  "What is 13-24 or 25-36 ?" which returned "loan is paid during months" as the answer.
\end{itemize}

\begin{table*}[t!]
\centering
\resizebox{10cm}{!}
{
    \begin{tabular}{lll}
\hline
\multicolumn{1}{c|}{\textbf{Phrases extracted}}                                                                & \multicolumn{1}{c|}{\textbf{Question}}                                                                               & \multicolumn{1}{c}{\textbf{Answer}} \\ \hline
\multicolumn{1}{l|}{\begin{tabular}[c]{@{}l@{}}borrowing capacity,  \\ available credit facility\end{tabular}} & \multicolumn{1}{l|}{\begin{tabular}[c]{@{}l@{}}What is borrowing capacity \\ on evolving credit loan ?\end{tabular}} & \$60,000 to \$93,000                  \\
\multicolumn{1}{l|}{\begin{tabular}[c]{@{}l@{}}borrowing capacity  \\ on revolving credit loan\end{tabular}}   & \multicolumn{1}{l|}{\begin{tabular}[c]{@{}l@{}}How much is available \\ credit facility ?\end{tabular}}              & \$33,000                            \\ \hline
                                                                                                               &                                                                                                                      &                                     \\ \hline
\multicolumn{1}{l|}{penalty of \%, loan balance}                                                               & \multicolumn{1}{l|}{What is 13-24 or 25-36 ?}                                                                        & loan is paid during months          \\
\multicolumn{1}{l|}{date of repayment}                                                                         & \multicolumn{1}{l|}{What is 2\% and 1\%}                                                                             & 2\% (Wrong Answer)                  \\ \hline
                                                                                                               &                                                                                                                      &                                     \\ \hline
\multicolumn{1}{l|}{lease liabilities, discount rate}                                                          & \multicolumn{1}{l|}{What is average lease term ?}                                                                    & 10.3 years                          \\ \cline{3-3} 
\multicolumn{1}{l|}{average lease term}                                                                        & \multicolumn{1}{l|}{What is discount rate ?}                                                                         & 2.00\%                              \\ \hline
\end{tabular}
}
\caption{Illustration of the baseline approach based on sentences in \S \ref{baseline} }\label{apptab2}
\end{table*}

Using the rules stated above, the entity and its associated noun phrases are identified. The last two columns of Table \ref{apptab2} show the questions which were generated and their responses from the MRC model. Inspired by the success of the pre-trained transformer model, we employ distilled BERT \cite{sanh2019distilbert} by Hugging Face \cite{wolf2020transformers} trained on SQuAD dataset \cite{rajpurkar2016squad} as the MRC model for our zero-shot question answering \footnote{Hugging Face's Model Link: \url{https://huggingface.co/transformers/v2.8.0/usage.html}.}.

\subsection{Baseline Method Code:}

\paragraph{sent\_parse:} This function takes a row as input, which has a column named "paragraph", and tokenizes the paragraph into sentences using the sent\_tokenize function from the nltk library. 
It then iterates through each sentence and checks if the value in the row is present in the sentence. 
If it is, the function returns the sentence. 
If not, the function does nothing.

\paragraph{sentence\_entity\_flair(sentence,entity, entity\_type):} This function takes a sentence, an entity, and an entity type as input. 
It first removes words between parentheses that do not contain digits, as well as any forward slashes. 
It then removes any brackets surrounding a dollar amount. 
The function then uses an entity\_tagger function to identify entities in the sentence, and iterates through each identified entity. 
Depending on the entity label and the entity type provided, the function checks if the entity matches the given entity. 
If it does, the function creates a list containing the entity, its label, and the original sentence, and returns it. 
If no matching entity is found, the function returns a list containing the original entity, a label of "none", and the original sentence.

\paragraph{preposition\_phrase\_extraction:} This function takes a text as input and uses the nlp function from the spacy library to parse the text. 
It then iterates through each token in the parsed text, and if the token is an adposition (preposition), it checks if its headword is a noun or pronoun. 
If it is, the function creates a phrase by appending any adjectives or compound dependencies of the head noun, the head noun itself, and any nouns or proper nouns to the right of the preposition, along with the preposition. 
The function then returns a list of all phrases found.

\paragraph{noun\_phrase\_extraction:} This function takes a text as input and uses the nlp function from the spacy library to parse the text. 
It then iterates through each token in the parsed text, and if the token is a noun or proper noun and its dependency is either "dobj," "pobj," "nsubj," or "nsubjpass," the function creates a phrase by appending any adjectives or compound dependencies of the noun and the noun itself. 
The function then returns a list of all phrases found.

\paragraph{phrase\_extraction:} This function takes a text as input and uses the entity\_tagger function to identify entities in the text. 
It then uses the preposition\_phrase\_extraction and noun\_phrase\_extraction functions to extract phrases from the text. 
For each extracted phrase, the function checks if it is present in any of the identified entities. 
If it is not, the function appends the phrase to a list of phrases to return. 
The function then returns the list of phrases.

\section{Other Results}

\begin{figure*}[h!]
	\centering
	\includegraphics[width=\linewidth, height= 4 cm]{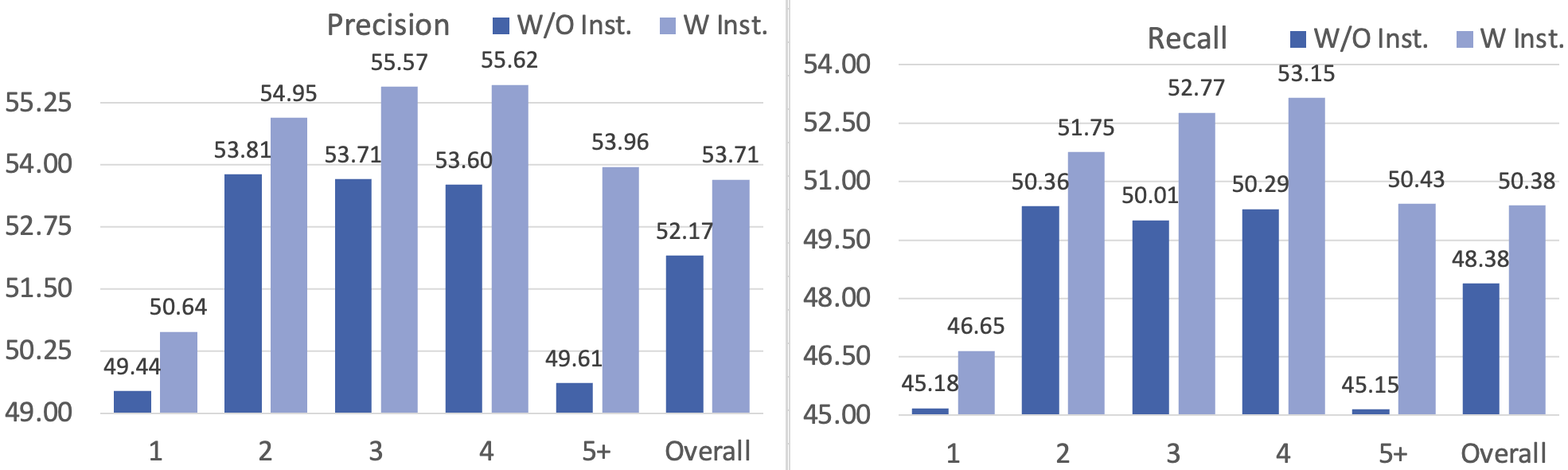}
	\caption{Illustrates the overall process flow of the proposed zero-shot open information extraction technique using question generation and reading comprehension.}
	\label{precision_recall}
\end{figure*}

Figure \ref{precision_recall} shows the precision and recall scores of instruction tuning Tk-Instruct on EDGAR10-Q. 
Similar trends are observed for precision and recall as for F1 score.

\end{document}